\documentclass[10pt]{amsart}
\usepackage{amssymb}
\usepackage[latin1]{inputenc}
\usepackage{amsmath}
\usepackage{hyperref}
\newtheorem{theorem}{Theorem}[section]
\newtheorem{lemma}[theorem]{Lemma}
\newtheorem{proposition}[theorem]{Proposition}

\newtheorem{definition}[theorem]{Definition}

\newtheorem{remark}[theorem]{Remark}
\numberwithin{equation}{section}
\newcommand{\vol}{V}
\newcommand{\volp}{v}

\begin{document}

\title[American Option Pricing]{Viscosity Solutions and American Option Pricing in a Stochastic Volatility Model of the Ornstein-Uhlenbeck Type}

\author{Alexandre F. Roch}
\thanks{* This research was supported in part by the NSERC, the IFM2
and the Fonds québécois de la recherche sur la nature et les
technologies.}
\address{Alexandre F. Roch: Center for Applied Mathematics,  Cornell University, Ithaca, NY 14853, USA}
\email{alexandre.f.roch@gmail.com}
\urladdr{http://www.cam.cornell.edu/$\sim$roch}


\begin{abstract}
In this paper, we study the valuation of American type derivatives
in the
 stochastic volatility model of Barndorff-Nielsen and Shephard \cite{BNS2001}. We characterize the value of such derivatives as
 the unique viscosity solution of an integral-partial differential
 equation when the payoff function satisfies a Lipschitz condition.
\end{abstract}

\maketitle

\section{Introduction}\label{introduction}

In their seminal paper, Barndorff-Nielsen and Shephard
\cite{BNS2001} introduced a model that has been shown to describe
particularly well financial assets for which log-returns have
heavy tail distributions and display long range dependence. In this model, the
volatility of the asset is described by an Ornstein-Uhlenbeck type process with
a pure jump Lévy process acting as the background driving process.
An empirical study was made in \cite{BNS2001} and showed from
exchange rate data that suitable distributions for the Lévy
process are the so-called generalized inverse gaussian
distributions from which well understood examples are the normal
inverse gaussian (studied in \cite{BN1998}) and the gamma
distribution.

The BNS model has been studied from different points of view.
 Benth et al. \cite{BKR2003} considered the problem of optimal
  portfolio selection.
Nicolato and Vernados \cite{NV2003} have studied European option
pricing and described the set of equivalent martingale measures
under this model. To evaluate these types of options, the authors
propose the transform-based method and a simple Monte Carlo
method.

In this paper, we consider the pricing of American options with the use of integral-partial differential equations
(IPDE). Although our technique can be simplified and used for
European options and certain path dependent options such as
barrier options (see \cite{CV2005} for a definition and examples),
we will mainly concentrate on American type derivatives which have
not been studied for this model. The main difficulty in this case
is the lack of Lipschitz continuity
of some of the coefficients of the IPDE.

The connection between viscosity solutions of IPDE's and Lévy
processes has been studied in the literature by various authors.
Pham \cite{P1998} considered a general stopping time problem of a
controlled jump diffusion processes. However, his results do not
apply here because the Lipschitz condition on the coefficients
is not satisfied in our current setting. Cont and Voltchkova
\cite{CV2005} studied barrier options and Barles et al.
\cite{BBP1997} established the connection between viscosity
solutions and backward stochastic differential equations. In these
papers, the stock price considered is modelled by a stochastic
differential equation with jumps driven by a Lévy process. The
main difference between the BNS model and these models is the
presence of stochastic volatility. However, we will see that the
lack of smoothness of the solution to our IPDE will also lead us
to consider the notion of viscosity solutions as presented in
\cite{CIL1992}.

The rest of the paper is organized as follows. In Section \ref{setup}, we
present the model and recall the results of Nicolato and Vernados
\cite{NV2003} regarding the set of equivalent martingale
measures. Section \ref{sectioncontinuity} is devoted to the continuity of the value function. In Section \ref{viscositysolutionssection}, we prove
that the value function is the viscosity solution of
the associated IPDE and the uniqueness of the solution is presented in Section \ref{Uniqueness}.

\section{Lévy Processes and the BNS Model}\label{setup}
Let $T>0.$ We consider
the stochastic volatility model of Barndorff-Nielsen and Shephard
\cite{BNS2001} for the price process of an asset, denoted by
$S=\{S_t\}_{0 \leq t \leq T}$ and defined on a filtered probability space
$(\mathrm{\Omega},
\mathcal{F},\{\mathcal{F}_t\}_{0 \leq t \leq T},\mathbb{P})$. We thus assume that the log-return
$X_t=\log(S_t)$ of the asset satisfies the following stochastic differential
equation:
\begin{eqnarray}
dX_{t} &=&(\mu +\beta \vol_t)dt+\sqrt{\vol_{t}} d
B_{t}+\rho dZ_{\lambda t}  \label{eq1} \end{eqnarray} with
\begin{eqnarray} d\vol_{t} &=&-\lambda \vol_{t} d t+d Z_{\lambda t} \label{eq2}
\end{eqnarray}
in which $\mu ,\beta \in \mathbb{R}$, $\lambda >0$ and $\rho \leq 0.$
 $B=\{B_{t}\}_{0 \leq t \leq T}$ is a Brownian motion and
$Z=\{Z_{t}\}_{0 \leq t \leq T}$ is the background driving  L{\'e}vy process
(BDLP) under the physical measure $\mathbb{P}$. In this model, $Z$
has no gaussian component and the increments are positive. $Z$ and $B$ are
assumed to be independent, and $\mathbb{F} =
\{\mathcal{F}_t\}_{0 \leq t \leq T}$ is the usual filtration generated by
the pair $(B,Z)$. The
positivity of the jumps of $Z$ insure that the process $\vol$
is always positive. We denote by $W$ the L{\'e}vy measure of $Z$.

Suppose $\mathbb{Q}$ is a probabilty measure equivalent to $\mathbb{P}$ under
which $S$ is a martingale. We are interested in
American-type derivatives of the form
\begin{eqnarray*}
U_t= {esssup}_{\tau \in \mathcal{T}_T, \tau \geq t}
\mathbf{E}_\mathbb{Q}[e^{-r(\tau-t)}h(X_\tau) | \mathcal{F}_t]
\end{eqnarray*} in which $h$ is the
payoff function and  $\mathcal{T}_T$ is the set
of all stopping times with values less or equal to $T$. Since
$\{X_t\}_{0\leq t \leq T}$ and $\{\vol_t\}_{0\leq t \leq T}$
are Markov processes, $U_t$ can be written as a function of
$(x,\volp,t)$, say
\begin{eqnarray*}U_t = u(x,\volp,t)= \sup_{\tau \in \mathcal{T}_{T-t}} \mathbf{E}_\mathbb{Q} \Big(e^{-r \tau}h(X^{x,\volp}_\tau) \Big) \end{eqnarray*} in which $(X^{x,\volp}_t)_{t \geq 0}$ is the process $X$ for which $X_0 = x$ and $\vol_0 = \volp.$ We also denote by  $(\vol^{\volp}_t)_{t \geq 0}$  the process $\vol$ starting at $\vol_0 = \volp$ at $t=0.$

\subsection{Equivalent Martingale Measures} \label{EMMsection}
 We start by summarizing the results of Nicolato
and Vernados \cite{NV2003} concerning the set of
equivalent martingale measures. In order to do so, we define the
set
\begin{eqnarray*}
\mathcal{Y}^{^{\prime }}= \left\{ y:[0,\infty )\rightarrow \lbrack
0,\infty )\ ; \ \int_{0}^{\infty }(\sqrt{y(x)}-1)^{2}w(x)dx<\infty
\right\}
\end{eqnarray*} and $\mathcal{M}'$ as the set of all
equivalent martingale measures $\mathbb{Q}$ such that  $Z$ is
still a L{\'e}vy process under $\mathbb{Q}$ independent of $B$,
possibly with a different marginal distribution.

As in \cite{NV2003}, we impose the following conditions on the
process $Z$:
\begin{enumerate}
\item[(C1)] The process $Z$ is given by the characteristic triplet
$(0,0,W)$ so that the cumulant transform is given by
\begin{eqnarray*}
\kappa (\theta )=\log \{\mathbf{E}[\exp (\theta
Z_{1})]\}=\int_{0}^{\infty }(e^{\theta z}-1)W(d z);
\end{eqnarray*} for values of $\theta$ for which this expression
is defined.
 \item[(C2)] $\widehat{\theta }=\sup \{\theta \in
\mathbb{R} \ |\ \kappa (\theta ) < \infty  \}>0$;
 \item[(C3)]
$\displaystyle \lim_{\theta \rightarrow \widehat{\theta }}\kappa
(\theta )=\infty$.
\end{enumerate}

\begin{remark}
Assumption (C2) implies that there exists $\widehat{\theta}_0>0$
such that
\begin{eqnarray*}
\int_{0}^{\infty }(e^{\widehat{\theta}_0 z}-1)W(d z) < \infty.
\end{eqnarray*} For $z>0$ and $n \geq 1$, we have
$ 0 < z^n \leq \frac{n!}{\widehat{\theta}_0^{n}}
(e^{\widehat{\theta}_0 z}-1), $ so that
$
\mu_n := \int_{0}^{\infty } z^n W(d z) < \infty.
$
Furthermore, Assumption (C2) is a sufficient condition for the
process $Z$ to have finite moments of all orders.
\end{remark}

The following theorem was proved in \cite{NV2003}.
\begin{theorem} \label{thEMM}
For all $\mathbb{Q} \mathcal{\in M}^{^{\prime }},$ there exists
$y\in \mathcal{Y} ^{^{\prime }}$ such that
\begin{eqnarray*}
dX_{t} &=& \left(r-\lambda \kappa ^{y}(\rho )-\frac{1}{2}\vol
_{t}\right)dt+\sqrt{\vol_{t}} dB_{t}^{\mathbb{Q}}+\rho dZ_{\lambda t}, \\
\text{in which }\kappa ^{y}(\theta ) &=&\int_{0}^{\infty }(e^{\theta
x}-1)y(x)w(x)dx,
\end{eqnarray*}
and $B_{t}^{\mathbb{Q}}=B_{t}-\int_{0}^{t} (\sqrt{\vol_{s}})^{-1}(r-\mu -(\beta +\frac{%
1}{2})\vol_{s}-\lambda \kappa ^{y}(\rho ))ds$ and $Z_{\lambda
t}$ are respectively a Brownian motion and a L{\'e}vy process
under $\mathbb{Q}$. $w^{y}(x)=y(x)w(x)$ is the L{\'e}vy density of
$Z_1$ under $\mathbb{Q}$ and $\kappa ^{y}(\theta )$ is the
cumulant function.
\end{theorem}

In the remaining part of this paper, all expectations will be with
respect to a chosen EMM $\mathbb{Q}$, unless specified otherwise,
and $W$ and $B$ will denote the associated L{\'e}vy measure and
the Brownian motion associated to $\mathbb{Q}$.

Let $\mathcal{O} = \mathbb{R}\times \mathbb{R}_+ \times [0,T)$ and
assume for a moment that $u$ is Lipschitz in $(x,\volp)$ and
\begin{eqnarray}\label{diff}
u \in C^{2,1,1}(\mathcal{O}),
\end{eqnarray}
that is $u$ is differentiable with respect to $\volp$ and $t$,
and twice differentiable with respect to $x$. We can then apply
Itô's formula to $U$ to find
\begin{eqnarray}
dU_t &=& (\frac{\partial u}{\partial t} + \mathcal{L}[u]) d t  +
\frac{\partial u}{\partial x} \sqrt{\vol_t} dB_t + d
\mathcal{V}_t, \label{ItoV}
\end{eqnarray} in which
\begin{eqnarray*}
\mathcal{L}[u] &=&(r -\frac{1}{2} \volp -\lambda
\kappa^y(\rho) + \lambda \rho \mu_1) \frac{\partial u}{\partial x}
- \lambda (\volp - \mu_1) \frac{\partial u}{\partial
\volp} +\frac{1}{2} \volp \frac{\partial^2 u}{\partial
x^2} \\
&& +\lambda \int_0^\infty (u(x+\rho z ,\volp +
z,t)-u(x_,\volp,t) - (\rho z \frac{\partial u}{\partial x} + z
\frac{\partial u}{\partial \volp}) ) W( d z).
\end{eqnarray*}
and $\mathcal{V}_t$ is the $\mathbb{Q}$-martingale given by
\begin{eqnarray*}
d \mathcal{V}_t &= & \int_0^\infty \Big(u(X_{t-}+ \rho
z,\vol_{t-} + z,t)-u(X_{t-},\vol_{t-},t)\\
&& \qquad \qquad -(\rho z \frac{\partial u}{\partial
x}(X_{t-},\vol_{t-},t) + z \frac{\partial u}{\partial
\volp}(X_{t-},\vol_{t-},t))\Big)
\widetilde{N}(dz,\lambda dt)\\
&&+\int_0^\infty \left(\rho z \frac{\partial u}{\partial
x}(X_{t-},\vol_{t-},t) + z \frac{\partial u}{\partial
\volp}(X_{t-},\vol_{t-},t)\right)
\widetilde{N}(dz,\lambda dt)
\end{eqnarray*} in which  $\widetilde{N}(dz,dt)=N(dz,dt)-W(dz)dt$ and $N(dz,dt)$ is the random measure of the
process $Z$. Since $\int_0^t \frac{\partial u}{\partial x}
\sqrt{\vol_t} dB_t$ is a $\mathbb{Q}$-martingale, if it can
be shown that $e^{-r t} U_t$ is also a martingale we can then
expect $u$ to satisfy the following integral-partial differential
equation (IPDE)
\begin{eqnarray*}
\frac{\partial u}{\partial t}(x,\volp,t) +
\mathcal{L}[u](x,\volp,t) - r u(x,\volp,t) = 0
\end{eqnarray*} if $ u(x,\volp,t) > h(x).$
 Otherwise $u(x,\volp,t) = h(x)$ and this IPDE can be written as
 \begin{eqnarray} \label{AmIPDE}
\max(\frac{\partial u}{\partial t}(x,\volp,t) +
\mathcal{L}[u](x,\volp,t) - r u(x,\volp,t),
h(x)-u(x,\volp,t)) = 0.
\end{eqnarray} It is clear also that the function satisfies
\begin{eqnarray}\label{AmIPDEboundarycond}
 u(x,\volp,t)=h(x) \mbox{ for $\volp=0$ or $t=T.$}
\end{eqnarray}

Condition \ref{diff} is in fact very restrictive and most of the
time not satisfied. Despite this problem, we will see that $u$ can
still be regarded as a solution of this equation in a weaker
sense.

\section{Continuity of the Value Function}\label{sectioncontinuity}

Recall the definition of the value of an American option with payoff $h$:
\begin{eqnarray}\label{Vam}
u(x,\volp,t)= \sup_{\tau \in \mathcal{T}_{T-t}}
\mathbf{E}\left(e^{-r \tau}h(X^{x,\volp}_\tau)\right).
\end{eqnarray} In the rest of this paper, we will assume that $h$ is positive and
satisfies the Lipschitz condition, in other words $\exists K>0$
such that $\forall (x_1,x_2) \in \mathbb{R}^2$
\begin{eqnarray}\label{Lipschitz}
|h(x_1)-h(x_2)|\leq K |x_1-x_2|.
\end{eqnarray}
For instance, the payoff function for an American put with strike $\tilde{X}>0$ is $h(x) =
\max(\tilde{X}-\exp(x),0)$ and satisfies this condition.

Our goal is to show that the function $u$ satisfies the IPDE
(\ref{AmIPDE}) in some weak sense. In order to give meaning to
this IPDE for a function $u$ that doesn't satisfy basic
differentiability conditions, we introduce the idea of viscosity
solutions following Crandall and Lions \cite{CL1983}. Let
$\mathcal{W}$ be the set of functions $f:\overline{\mathcal{O}}
\rightarrow \mathbb{R}$ that satisfy
\begin{eqnarray*}
 \sup_{(x,\volp),(x',\volp') \in
\mathbb{R}\times \mathbb{R}_+ }
\frac{|f(x,\volp,t)-f(x',\volp',t)|}{1+|x-x'|+|\volp-\volp'|}<
\infty \quad \forall t \in [0,T] .\end{eqnarray*}

\begin{definition}\label{super}
The function $u \in C^0(\overline{\mathcal{O}}) \cap \mathcal{W}$
is a viscosity subsolution (supersolution) of
(\ref{AmIPDE})-(\ref{AmIPDEboundarycond}) if $\forall
(x,\volp,t) \in \mathcal{O}$ and $\forall \psi \in \mathcal{W}
\cap C^{2,1,1}(\mathcal{O})$ such that
\begin{enumerate}
\item[(i)] $\psi(x,\volp,t)=u(x,\volp,t)$ and
\item[(ii)]$\forall (x',\volp',t')\in \mathcal{O}
\quad\psi(x',\volp',t') \geq u(x',\volp',t') \quad
(\leq),$
\end{enumerate} then
\begin{eqnarray} \nonumber \max \Big( \frac{\partial \psi(x,\volp,t)}{\partial t} +
\mathcal{L}[\psi](x,\volp,t) - r \psi(x,\volp,t) \: ;
\\ \label{EIDPASX} h(x)-u(x,\volp,t) \Big) &\geq&  0
\quad (\leq),
\end{eqnarray} \begin{eqnarray}\label{EIDPASXC}
 \mbox{ and } u(x,\volp,t)&=&h(x) \mbox{ for $\volp=0$ or $t=T.$}
\end{eqnarray}  The function $u$ is a viscosity solution if
it is both subsolution and supersolution.
\end{definition}

\begin{remark}
As noted in \cite{CV2005} (p.317) the condition $\psi \in
\mathcal{W}$ is sufficient to have a well defined integral term in
$\mathcal{L}[\psi].$ In fact if $\psi \in \mathcal{W}\cap
C^{2,1,1}$ then \\
 $\int_0^\infty \Big(\psi(x +\rho z,
\volp+z,t) - \psi(x,\volp,t) - (\rho z \frac{\partial
\psi}{\partial x}(x,\volp,t) + z \frac{\partial \psi}{\partial \volp}(x,\volp,t))
\Big) W(d z)$
\begin{eqnarray*} & \leq& \int_{z<\eta} C z^2 W(d
z) + \int_\eta^\infty C (1+|z|) W(d z) < \infty
\end{eqnarray*} for any $\eta > 0$.
\end{remark}

An important property of viscosity solutions is the continuity of the function.
It is the content of the following proposition.

\begin{proposition} \label{continuite}
When $h$ satisfies the Lipschitz condition (\ref{Lipschitz}), the
function $u$ is continuous and in $\mathcal{W}$.
\end{proposition}
\proof
 In this proof, we will assume for simplicity that $r=0$.
The generalization to $r>0$ is straightforward.  Throughout, $C$
is a positive constant that can change from line to line.

We start by showing the continuity of $u$ with respect to
$(x,\volp)$, uniformly in $t$.  We have the following representation of the volatility process:
\begin{eqnarray*}
\vol_t^{\volp} &=& \volp e^{-\lambda t} +
\int_0^t e^{-\lambda s} d
Z_{\lambda s}.
\end{eqnarray*}
  We define
the integrated variance process started with $\vol_0 = \volp$ by
\begin{eqnarray*}
\vol_{t}^{\volp,*}&=&\int_{0}^{t} \vol^\volp_s d s.
\end{eqnarray*} By Equation (\ref{eq2}), we find that
\begin{eqnarray*}\vol^{\volp}_t d t&=& \frac{1}{\lambda} (- d \vol^{\volp}_t + d Z_{\lambda
t})\end{eqnarray*} so that we have the following representation of
the integrated variance process
\begin{eqnarray}
\vol_{t}^{\volp,*} &=& \frac{1}{\lambda}(\volp - \vol_t^\volp) + \frac{1}{\lambda} \int_0^t d Z_{\lambda s}\\
&=  &\volp \varepsilon(t) + \int_0^t \varepsilon(s) d Z_{\lambda s}. \label{integratedvol}
\end{eqnarray} in which $\varepsilon(t) = \frac{1-e^{-\lambda t}}{\lambda}$.

 We also have the following identities:
\begin{eqnarray*}
X_t^{x,\volp} &=& x  -\lambda \kappa(\rho) t -
\frac{1}{2} \vol_{t}^{\volp, *} + \int_0^t
\sqrt{\vol_s^{\volp}} d B_s + \rho Z_{\lambda t}, \\
\Delta \vol_t &:=& \vol_t^{\volp'} -
\vol_t^{\volp} = \Delta \volp
e^{-\lambda t}, \\
\Delta \vol_t^* &:=& \vol_{t}^{\volp',*} -
\vol_{t}^{\volp,*} = \Delta
\volp \varepsilon(t), \\
\Delta X_t &:=& X_t^{x',\volp'} - X_t^{x,\volp} = x'-x
- \frac{1}{2} \Delta \volp \varepsilon(t) + \int_0^t \left(
\sqrt{\vol_s^{\volp'}}-\sqrt{\vol_s^{\volp}} \right) d B_s,\\
&:=& \Delta x - \frac{1}{2} \Delta \volp \varepsilon(t) +
M_{t}^{\volp,\volp'},
\end{eqnarray*}
with  $\Delta
x = x'-x$ and $\Delta \volp = \volp'-\volp$.

Using the Lipschitz condition on $h$, we obtain
\begin{eqnarray*}
\left|u(x',\volp',t)-u(x,\volp,t)\right|&=&
\left|\sup_{\tau \in \mathcal{T}_{T-t}} \mathbf{E}
h(X_{\tau}^{x',\volp'})-\sup_{\tau \in \mathcal{T}_{T-t}}
\mathbf{E} h(X_{\tau}^{x,\volp}) \right|\\
&\leq& \sup_{\tau \in \mathcal{T}_{T-t}} \mathbf{E} \left|
h(X_{\tau}^{x',\volp'})-h(X_{\tau}^{x,\volp}) \right| \\
&\leq& C \sup_{\tau \in \mathcal{T}_{T-t}} \mathbf{E} \left|
X_{\tau}^{x',\volp'}-X_{\tau}^{x,\volp} \right|.
\end{eqnarray*}

Then,
\begin{eqnarray*}
\left|u(x',\volp',t)-u(x,\volp,t)\right| &\leq&  C \Big(
|\Delta x| +|\Delta \volp|+ \sup_{\tau \in \mathcal{T}_{T-t}}
\mathbf{E}|M_{\tau}^{\volp,\volp'}|  \Big).
\end{eqnarray*}

Letting $\mathcal{G} = \sigma(\{Z_s\}_{0\leq s \leq T})$, the
$\sigma$-field generated by the BDLP $Z$ up to the maturity $T$,
we find that $\{M_{t}^{\volp,\volp'}\}_{t \geq 0}$ is a
$\mathcal{G} \vee \mathcal{F}_t$-martingale. Thus, $\{|M_{t}^{\volp,\volp'}| \}_{t \geq 0}$
is a $\mathcal{G} \vee \mathcal{F}_t$-submartingale and Doob's theorem applies. In other words,
\begin{eqnarray*}
\sup_{\tau \in \mathcal{T}_{T-t}} \mathbf{E}  |M_{\tau}^{\volp,\volp'}|
&\leq& \mathbf{E} \Big( \sup_{\tau \in
\mathcal{T}_{T-t}} \mathbf{E} \Big( |M_{\tau}^{\volp,\volp'}| \Big| \mathcal{G} \Big) \Big)\\ &\leq&
\mathbf{E} \left( \mathbf{E}\left( |M_{T-t}^{\volp,\volp'}|\Big| \mathcal{G} \right) \right)  \leq
\sqrt{\mathbf{E}\Big( \mathbf{E}\Big((M_{T-t}^{\volp,\volp'})^2 \Big| \mathcal{G} \Big) \Big)}.
\end{eqnarray*}
Also, \\
$\mathbf{E}\Big( (M_{T-t}^{\volp,\volp'})^2 \Big| \mathcal{G}
\Big)$
\begin{eqnarray*}
 &=& \int_0^{T-t} (\vol_s^{\volp'} - 2
\sqrt{\vol_s^{\volp'}\vol_s^{\volp}} +
\vol_s^\volp) d s
\\
&=& \int_0^{T-t} \Delta \volp e^{-\lambda s} d s + 2 \int_0^{T-t}
\vol_s^{\volp} - \sqrt{{(\vol_s^{\volp})}^2 +
\vol_s^{\volp} \Delta \volp
e^{-\lambda s}} d s \\
&=&\int_0^{T-t} \Delta \volp e^{-\lambda s} d s + 2 \int_0^{T-t}
\frac{-\vol_s^{\volp} \Delta \volp e^{-\lambda
s}}{\vol_s^{\volp} +
\sqrt{{(\vol_s^{\volp})}^2 + \vol_s^{\volp}
\Delta \volp e^{-\lambda
s}}} d s\\
&\leq& \int_0^{T-t}  3 | \Delta \volp | e^{-\lambda s} d s  = 3 |\Delta \volp| \varepsilon(T-t) \leq 3 |\Delta \volp| T. \\
\end{eqnarray*}

And thus we proved the continuity of $u$ in
 $(x,\volp)$ uniformly in $t$ since
 \begin{eqnarray*}\left|u(x',\volp',t)-u(x,\volp,t)\right| \leq
    C \left( |\Delta x| +|\Delta \volp|+ \sqrt{|\Delta \volp|}\right). \end{eqnarray*} In particular  $u \in \mathcal{W}$
because of the following inequality
\begin{eqnarray*}\left|u(x',\volp',t)-u(x,\volp,t)\right| &\leq&  C \left( |\Delta x|
+|\Delta \volp|+ \sqrt{|\Delta \volp|} \right) \\ &\leq&
2C(1+ |\Delta x| +|\Delta \volp|).\end{eqnarray*}

The next step of the proof is to show
\begin{eqnarray*}
\mathbf{E} \sup_{t \leq s \leq t'} \left| X^{x,\volp}_s-X^{x,\volp}_t \right|
\rightarrow 0 \quad \mbox{and} \quad \mathbf{E} \sup_{t \leq s\leq
t'} \left| \vol^\volp_s-\vol^\volp_t \right| \rightarrow 0
\end{eqnarray*} as $|t-t'| \rightarrow 0.$
This is easily obtained by first observing that\\
$\mathbf{E} \sup_{t \leq s \leq t'} \left| X^{x,\volp}_s-X^{x,\volp}_t \right|$
\begin{eqnarray*}
 &\leq&
\frac{1}{2} \mathbf{E} \sup_{t \leq s \leq t'} \left|
\vol_{s}^{\volp,*}-\vol_{t}^{\volp,*} \right| + \mathbf{E} \sup_{t \leq s \leq t'}
\left| \int_t^s \sqrt{\vol^\volp_{y}} d B_{y} \right| + \rho \mathbf{E}
\sup_{t \leq s \leq t'} \left| Z_{\lambda s}
-Z_{\lambda t} \right|\\
&\leq& \frac{1}{2} \volp |\varepsilon(t')-\varepsilon(t)| + \mathbf{E}
\left| \int_t^{t'} e^{-\lambda(t'-s)} d Z_{\lambda s} \right|
\\
&&\quad  + C \sqrt{\mathbf{E} \left| \vol_{t'}^{\volp,*}-\vol_{t}^{\volp,*} \right|} +
\rho \mathbf{E}
\left| Z_{\lambda t'}-Z_{\lambda t} \right|\\
&\leq & \frac{1}{2} \volp |\varepsilon(t')-\varepsilon(t)| + (1+\rho)
\mathbf{E} \left| Z_{\lambda t'} - Z_{\lambda t}\right|\\ && \quad
+  \sqrt{\frac{1}{2} \volp |\varepsilon(t')-\varepsilon(t)|
 + \mathbf{E} \left|
Z_{\lambda t'} - Z_{\lambda t}\right| }.
\end{eqnarray*} As for the process $\vol$,
\begin{eqnarray*}
\mathbf{E}\sup_{t \leq s \leq t'}  \left| \vol^\volp_s-\vol^\volp_t
\right| &\leq& |1-e^{-\lambda (t'-t)} | \: \mathbf{E} |
\vol^\volp_{t} | + \mathbf{E} \left| \int_t^{t'} e^{-\lambda(t'-s)} d
Z_{\lambda s} \right| \\
&\leq& |1-e^{-\lambda (t'-t)} | \: \mathbf{E} | \vol^\volp_{t} | +
\mathbf{E} \left| Z_{\lambda t'} - Z_{\lambda t} \right|,
\end{eqnarray*} Since $\vol^\volp_t \leq \volp + Z_{\lambda T}$ for all $t \leq T$,
\begin{eqnarray*}
\mathbf{E}\sup_{t \leq s \leq t'} \left| \vol^\volp_s-\vol^\volp_t
\right| \leq C (\volp+ \: \mathbf{E} Z_{\lambda T}) |t'-t| +
\mathbf{E} \left| Z_{\lambda t'} - Z_{\lambda t} \right|,
\end{eqnarray*}
and we need to show that $\mathbf{E} \left| Z_{\lambda t'} -
Z_{\lambda t} \right| \rightarrow 0$ when $|t'-t|\rightarrow 0.$

We mentioned earlier that condition (C2) implies that the moments
of $Z_t$ are finite for all orders. Thus $Z$ is uniformly
integrable. Since $Z$ is also continuous in probability, it is
continuous in
 $\mathcal{L}_1$ and the conclusion follows.

Let's now show continuity with respect to time. Let $0 \leq t \leq
t' \leq T$. Take $\tau \in \mathcal{T}_{T-t}$ and define $\tau' =\tau \wedge (T-t').$ Then,

$\mathbf{E}\left( e^{-r \tau}
h(X_\tau^{x, \volp})\right)$
\begin{eqnarray*}
 & = & \mathbf{E}\left(
e^{-r \tau'} h(X^{x, \volp}_{\tau'})\right) + \mathbf{E}\left( e^{-r \tau} h(X^{x, \volp}_\tau) -e^{-r \tau'} h(X^{x, \volp}_{\tau'})
\right) \\
& \leq & u(x, \volp, t') + \mathbf{E}\left( e^{-r \tau} h(X^{x, \volp}_\tau) -e^{-r \tau'} h(X^{x, \volp}_{\tau'})
\right).
\end{eqnarray*} From this inequality, we readily find that
\begin{eqnarray*}
| u(x, \volp, t') - u(x, \volp, t) | &\leq&
\mathbf{E} \sup_{T-t' \leq s \leq T-t} | X^{x, \volp}_s -
X^{x, \volp}_{T-t'}|
\end{eqnarray*} which converges to zero as $|t-t'|\to 0.$

Global continuity follows from the following inequality
\begin{eqnarray*}
\left|u(x', \volp', t')\!-\! u(x, \volp, t)\right|  \le
\left|u(x', \volp', t')\!- \!u(x, \volp, t')\right| +
\left|u(x, \volp, t')\!-\! u(x, \volp, t)\right|
\end{eqnarray*}
and the fact that the first bound is independent of $t'$. \qed

\section{Viscosity Solutions}
\label{viscositysolutionssection}

This section is devoted to the viscosity solution property of
the value function $u$.
In order to prove that $u$ is a viscosity solution of
(\ref{AmIPDE}), we need the following dynamic programming
principle. It is a consequence of the martingale property of the
Snell envelope stopped before its optimal stopping time and it is
the key property needed in the proof of the subsolution property.

\begin{lemma} \label{PPD}
Let $\epsilon>0$, $(x,\volp,t)\in \mathcal{O}$ and define the
stopping time
\begin{eqnarray*}\tau^\epsilon = \inf \{0 \leq s \leq T-t\:
|\: e^{-r s}u(X^{x,\volp}_s,\vol^{\volp}_s,t+s)-
\epsilon \leq e^{-r s}h(X^{x,\volp}_{s}) \}. \end{eqnarray*}
Then,
\begin{eqnarray} \label{PPDeq}
u(x,\volp,t)= \mathbf{E}[ e^{-r
 \tau^\epsilon}u(X^{x,\volp}_{\tau^\epsilon},\vol^{\volp}_{\tau^\epsilon},t+\tau^\epsilon)].
\end{eqnarray}
\end{lemma}
\proof For some constant $C$, we have that \begin{eqnarray*} n
\mathbf{E} \left(|\mathbf{1}_{\{h(X_\tau) \geq n\}}
h(X_{\tau})|\right)
 \leq
\mathbf{E}\left( h(X_{\tau})^2\right) \leq C + C \mathbf{E}\left(
X_{\tau}^2\right) \end{eqnarray*} for all $\tau \in
\mathcal{T}_{T}$. We know that $X_\tau = X_0 + r \tau +
\vol_{\tau}^* + \int_0^\tau \sqrt{\vol_s} d B_s + \rho
Z_{\lambda \tau}$ and that $0 \leq \vol_{\tau}^* \leq
\vol_{T}^* \leq \frac{1}{\lambda} (\vol_0 + Z_{\lambda
T})$ from Equation (\ref{integratedvol}). As a result, $X_\tau^2
\leq 4 (X_0 +r T)^2 + 4 \frac{1}{\lambda^2} (\vol_0 +
Z_{\lambda T})^2 + 4 \left( \int_0^\tau \sqrt{\vol_s} d B_s
\right)^2 + 4 \rho^2 Z_{\lambda T}^2 \leq C + C Z_{\lambda T}^2 +
C \left( \int_0^\tau \sqrt{\vol_s} d B_s \right)^2 $ for some
constant $C$ large enough. Hence $\mathbf{E} X_\tau^2 \leq C + C
\mathbf{E} Z_{\lambda T}^2 + C \mathbf{E} \vol_{T}^* <
\infty$ for all $\tau \in \mathcal{T}_T.$ As a consequence, $
\sup_{\tau \in \mathcal{T}_T} \mathbf{E}
\left(|\mathbf{1}_{\{h(X_\tau) \geq n\}} h(X_{\tau})|\right)$
converges to $0$ as $n$ grows to infinity, i.e. the collection
$\{e^{-r \tau }h(X_\tau) : \tau \in \mathcal{T}_T\}$ is uniformly integrable.
Hence we find that the process $(e^{- r s}h(X_{s}))_{0\leq s
\leq T}$ is of Class D and we can apply the results of
\cite{M1978} to get the result.
 \qed

The proof of the solution property of $u$ makes use of the
following lemma.
\begin{lemma}\label{lemmaprobatau} Let $t \leq T$ and $\epsilon >0.$ Suppose $u(x,\volp,t)-h(x)>\epsilon$. Then
$\mathbb{Q}( \tau^\epsilon < s) \rightarrow 0$ when $s
\rightarrow 0.$
\end{lemma}

\proof  Let  $\eta>\epsilon$ such that $\eta<u(x,\volp,t)-h(x)$.

 First we show that
 $e^{-r \tau^\epsilon} u(X^{x,\volp}_{\tau^{\epsilon}},\vol^{\volp}_{\tau{^\epsilon}},\tau{^\epsilon})
-e^{-r \tau^\epsilon} h(X^{x,\volp}_{\tau^{\epsilon}}) \leq \epsilon$ almost
surely. For some sequence $s_n \downarrow
\tau^\epsilon$,
 $e^{-r s_n} u(X^{x,\volp}_{s_n},\vol^\volp_{s_n},{s_n})\leq e^{-r s_n} h(X^{x,\volp}_{s_n}) +\epsilon$ for $n$ large enough. In this case,
since $(X^{x,\volp}_{s_n},\vol^\volp_{s_n})$ converges  to
$(X^{x,\volp}_{\tau^\epsilon},\vol^\volp_{\tau^\epsilon})$ in $\mathcal{L}^1$,
we can take a subsequence if necessary and find that
$|u(X^{x,\volp}_{s_n},\vol^\volp_{s_n},s_n)
-u(X^{x,\volp}_{\tau^\epsilon},\vol^\volp_{\tau^\epsilon},\tau_\epsilon)|\rightarrow 0
$ and $h(X^{x,\volp}_{s_n }) \rightarrow
h(X^{x,\volp}_{\tau^\epsilon})$ a.s. with $n\rightarrow \infty$. Taking the limit, we find
\begin{eqnarray*}
e^{-r \tau_\epsilon} u(X^{x,\volp}_{\tau^\epsilon},\vol^\volp_{\tau^\epsilon},\tau^\epsilon)& =&
\lim_{n\rightarrow \infty}
e^{-r s_n} u(X^{x,\volp}_{s_n},\vol^\volp_{s_n},s_n)\\
& \leq & \lim_{n \rightarrow \infty} e^{-r s_n} h(X^{x, \volp}_{s_n}) + \epsilon
\\& =&
e^{-r \tau_\epsilon} h(X^{x,\volp}_{\tau^\epsilon})+\epsilon \quad \mbox{a.s.}
\end{eqnarray*}

Since $u$ is continuous with respect to $t$, we find that  $\eta<e^{-r s} u(x,\volp,t+s)-e^{-r s} h(x)$ for $s$ small enough.
Then, for $s$ small enough,\\ $\mathbb{Q}( \tau^\epsilon < s)$
\begin{eqnarray*}
 &\leq& \mathbb{Q}\Big(
e^{-r \tau^\epsilon}(u(x,\volp,\tau^\epsilon) - h(x)) +
e^{-r \tau^\epsilon} (h(X^{x,\volp}_{\tau^{\epsilon}})- u(X^{x,\volp}_{\tau^{\epsilon}},\vol^{\volp}_{\tau{^\epsilon}},\tau^\epsilon))
 > \eta-\epsilon  \Big)\\
& \leq & \mathbb{Q}\Big( e^{-r \tau^\epsilon} \Big|  u(x,\volp,\tau^\epsilon) -
 u(X^{x,\volp}_{\tau^{\epsilon}},\vol^{\volp}_{\tau{^\epsilon}},\tau^\epsilon)\Big|+ e^{-r \tau^\epsilon} \Big|h(X^{x,\volp}_{\tau^{\epsilon}})-h(x)
\Big| > \eta-\epsilon \Big) \\
& \leq & \mathbb{Q}\Big( \Big|
\vol^{\volp}_{\tau^{\epsilon}} - \volp \Big|>
\delta_2 \Big)+\mathbb{Q}\Big(\Big|X^{x,\volp}_{\tau^{\epsilon}}- x
\Big|
> \delta_3 \Big)
\end{eqnarray*}  for some constants $\delta_2>0$ and $\delta_3>0$. By the continuity in probability of the processes $X$ et $\vol$, we know that this expression goes to zero when $s \to 0.$
\qed

We can now show that $u$ is a viscosity solution.
\begin{theorem}\label{viscoproperty}
When $h$ satisfies the Lipschitz condition (\ref{Lipschitz}), $u$
is a viscosity solution of IPDE (\ref{AmIPDE}).
\end{theorem}
\proof We  already know that $u$ is continuous and in
$\mathcal{W}$.

Let's start by showing that $u$ is a supersolution of
(\ref{AmIPDE}). Let $(x,\volp,t)\in \mathcal{O}$ and $\psi$
satisfy the conditions given in the above definition of
supersolutions. By
definition, for any $\Delta t>0,$
\begin{eqnarray*}
0 & \geq  &  e^{-r \Delta t} \mathbf{E} \left(
u(X^{x,\volp}_{\Delta t},\vol^\volp_{\Delta t},t+\Delta t)
\right)-u(x,\volp,t)\\
& \geq & \mathbf{E} \left(e^{-r \Delta t}
\psi(X^{x,\volp}_{\Delta t},\vol^\volp_{\Delta t},t+\Delta t)
-\psi(X^{x,\volp}_{0},\vol^\volp_{0},t) \right) \\
&= & \mathbf{E} \left( \int_0^{\Delta t}
e^{-r s} (-r  \psi +  \frac{\partial \psi}{\partial t}
+ \mathcal{L}[\psi] )(X^{x,\volp}_{s},\vol^\volp_{s},t+s) ds \right. \\
&& \quad + \left. \int_{0}^{\Delta t} \frac{\partial
\psi}{\partial x}(X^{x,\volp}_{s},\vol^\volp_{s},t+s) e^{-r s} \sqrt{\vol^\volp_s} dB_s  +
\Psi^{x,\volp}_{\Delta t}-\Psi^{x,\volp}_0  \right),
\end{eqnarray*}
in which $\Psi^{x,\volp}$ is the martingale defined by
\begin{eqnarray*}
d \Psi^{x,\volp}_s &= & e^{-r s} \int_0^\infty \Big(\psi(X^{x,\volp}_{s-}+ \rho
z,\vol^\volp_{s-} + z,t+s)-\psi(X^{x,\volp}_{s-},\vol^\volp_{s-},t+s)\\
&& \qquad \qquad - z \left(\rho  \frac{\partial \psi}{\partial
x} +  \frac{\partial \psi}{\partial
\volp}\right)(X^{x,\volp}_{s-},\vol^\volp_{s-},t+s)\Big)
\widetilde{N}(dz,\lambda ds)\\
&&+e^{-r s} \int_0^\infty z \left(\rho  \frac{\partial
\psi}{\partial x} +  \frac{\partial
\psi}{\partial \volp}\right)(X^{x,\volp}_{s-},\vol^\volp_{s-},t+s)
\widetilde{N}(dz,\lambda ds).\end{eqnarray*} Since
\begin{eqnarray*}
\int_{0}^{\Delta t} \frac{\partial \psi}{\partial x}(X^{x,\volp}_{s},\vol^\volp_{s},t+s) e^{-r s}
\sqrt{\vol^\volp_s} dB_s
\end{eqnarray*} is also a martingale, we have the following inequality
\begin{eqnarray*}0 \geq \int_0^{\Delta t} \mathbf{E} \left(
e^{-r s} (-r  \psi +  \frac{\partial \psi}{\partial t}
+ \mathcal{L}[\psi] )(X^{x,\volp}_{s},\vol^\volp_{s},t+s) \right)ds,\end{eqnarray*} in other words, dividing by $\Delta
 t$ and taking the limit as $\Delta t \rightarrow 0$
\begin{eqnarray*}
0 \geq -r \psi(x,\volp,t) +
 \frac{\partial \psi}{\partial t}(x,\volp,t) +\mathcal{L}[\psi](x,\volp,t). \end{eqnarray*} Since,
 by definition, $u(x,\volp,t) \geq h(x)$,
 $u$ satisfies Equation (\ref{AmIPDE}).
To prove that $u$ is a viscosity subsolution of (\ref{AmIPDE}),
let $(x,\volp,t)\in \mathcal{O}$ and $\psi$ satisfy the
conditions of the above definition for subsolutions. If
$u(x,\volp,t)= h(x)$, the inequality (\ref{EIDPASX}) is
satisfied. Otherwise, let
\begin{eqnarray*}0<\epsilon < u(x,\volp,t)-h(x).\end{eqnarray*}
We know from Lemma \ref{PPD} that \begin{eqnarray} 0 &=&
\mathbf{E}\left( e^{-r
 (\Delta t \wedge\tau^\epsilon)}u(X^{x,\volp}_{\Delta t\wedge\tau^\epsilon},\vol^\volp_{\Delta t \wedge\tau^\epsilon},t+{(\Delta t \wedge\tau^\epsilon)})
\right)- u(x,\volp,{t}) \nonumber \\ &\leq&
\mathbf{E}\left(e^{-r
(\Delta t \wedge\tau^\epsilon)}
\psi(X^{x,\volp}_{\Delta t \wedge\tau^\epsilon},\vol^\volp_{\Delta t \wedge\tau^\epsilon},t+{(\Delta t \wedge\tau^\epsilon)})
\right)- \psi(x,\volp,t) \nonumber \\
&=&\mathbf{E}\left(
\int_{0}^{{\Delta t \wedge\tau^\epsilon}}  e^{-r s} (-r \psi +
 \frac{\partial \psi}{\partial t} +\mathcal{L}[\psi]) (X^{x,\volp}_{s},\vol^\volp_{s},t+s)  d s  \right)
 \label{ineq}
 \end{eqnarray} for any $\Delta t > 0$. Knowing that $\mathbb{Q}(\tau^\epsilon<\Delta t)\rightarrow
 0$ when $\Delta t\rightarrow 0$ by Lemma \ref{lemmaprobatau},
 dividing the preceding inequality by $\Delta t$ and taking the limit to
 $0$, we get the desired result by Lebesgue's dominated convergence theorem.
 \qed

\section{Comparison Principles and Uniqueness of the Solution}\label{Uniqueness}

In this section, we prove a comparison result from which we obtain the uniqueness of the solution of the IPDE.
 In proving comparison results for viscosity solutions,  the notion
of parabolic superjet and subjet as defined in  Crandall et al.
\cite{CL1983} is
particularly useful. Setting $y=(x,\volp)$, we define the
parabolic superjet and its closure by
\begin{eqnarray*} \mathcal{J}^{2,+}u(y,t) &=& \Big\{(p,q,A) \in
\mathbb{R} \times \mathbb{R}^2 \times \mathcal{S}_2 \quad \mbox{ such that } \quad
u(y',t')-u(y,t) \leq \\&& \quad p \:(t'-t) + q  \cdot (y'-y) +
\frac{1}{2} (y'-y)^T \cdot A \cdot (y'-y) \\&&\quad+
o(|t'-t|+|y'-y|^2) \quad \mbox{as}\quad (t',y')
\rightarrow (t,y)\Big\}\\
\overline{\mathcal{J}}^{2,+}u(y,t) &= & \Big\{(p,q,A) =
\lim_{n\rightarrow \infty} (p_n,q_n,A_n)  \quad \mbox{such that}
\quad  \\&& \quad (p_n,q_n,A_n) \in \mathcal{J}^{2,+}u(y_n,t_n)
\quad \mbox{and} \quad (y_n,t_n) \rightarrow (y,t) \Big\}.
\end{eqnarray*} The subjet and its closure are then defined similarly
by
\begin{eqnarray*}
\mathcal{J}^{2,-}u(y,t)&=&- \mathcal{J}^{2,+}(-u)(y,t) \quad
\mbox{and} \\
 \overline{\mathcal{J}}^{2,-}u(y,t)&=&-
 \overline{\mathcal{J}}^{2,+}(-u)(y,t).
\end{eqnarray*}

We then have the following lemma which is essentially proved in
\cite{BBP1997} (lemma 3.3).
\begin{lemma}
If the  function $u \in C^0(\mathbb{R}\times \mathbb{R}_+ \times
[0,T])$ is a viscosity subsolution (resp. supersolution) of
(\ref{AmIPDE}) then $\forall (x,\volp,t) \in \mathbb{R} \times
\mathbb{R}_+ \times [0,T)$ and $\forall (p,q,A) \in
\overline{\mathcal{J}}^{2,+}u(x,\volp,t)$ (resp.
$\overline{\mathcal{J}}^{2,-}u(x,\volp,t)$)
\begin{eqnarray} \label{EIDPASXsuperjet}\max ( p +
\mathcal{L}_\xi^{q,A}[u,\psi](x,\volp,t) - r u(x,\volp,t)
\: ; \: h(x)-u(x,\volp,t) ) \geq  0 \quad (\leq),
\end{eqnarray} in which
\begin{eqnarray*} \mathcal{L}_\xi^{q,A}[u,\psi](x,\volp,t) = (r\!-\!\frac{1}{2} \volp\! -\! \lambda \kappa^y(\rho)\! +\! \lambda
\rho \mu^\eta) q^{(1)}\!  -\! \lambda (\volp\!-\!\mu^\eta)
q^{(2)} +\frac{1}{2} \volp A_{11}
\end{eqnarray*}
\begin{eqnarray*} + \lambda \int_0^{\xi} (\psi(x+\rho z,\volp + z,t)-\psi(x,
\volp, t ) - (\rho z \frac{\partial \psi}{\partial x} +  z
\frac{\partial \psi}{\partial \volp})) W(d z)
\end{eqnarray*} \begin{eqnarray*}+ \lambda \int_\xi^{\infty} (u(x+\rho z,\volp + z,t)-u(x,
\volp, t ) - (\rho z \frac{\partial \psi}{\partial x} + z
\frac{\partial \psi}{\partial \volp})) W(d z)\end{eqnarray*}
for some $\psi \in C^{2,1,1}$ and $0<\xi<1$.
\end{lemma}

 Pham
\cite{P1998} obtains the uniqueness of the solution when the coefficients of $\mathcal{L}$ satisfy Lipschitz
conditions on $\mathbb{R}^2 \times [0,T].$ For $\delta >0$,
define $\mathcal{O}^\delta =
\mathbb{R}\times (\delta, \infty) \times[0,T)$. Then, the coefficients of $\mathcal{L}$ satisfy the Lipshitz conditions on $\mathcal{O}^\delta$ and using the ideas of uniqueness proofs in the literature we can show a comparison principle on $\mathcal{O}^\delta.$ This result will then be used to show the uniqueness of the solution on $\mathcal{O}.$

\begin{theorem}\label{thComparison} Let $\epsilon \geq 0$, and $u_1$ be a subsolution and $u_2$ a supersolution of (\ref{AmIPDE}) on $\mathcal{O}^\delta$ such that
$$ u_1(x,\volp,t) \leq u_2(x,\volp,t)+\epsilon$$ for $t=T$ or $\volp = \delta.$ Then
$ u_1(x, \volp,t) \leq u_2(x, \volp,t) + \epsilon e^{r(T-t)}$ for all  $(x, \volp,t) \in \mathcal{O}^\delta$.
\end{theorem}

\begin{proof}
An IPDE of the form $\frac{\partial
\psi(x,\vartheta,t)}{\partial t} +
\mathcal{L}[\psi](x,\vartheta,t) - r \psi(x,\vartheta,t) = 0$ for
$(x,\vartheta,t) \in \mathcal{O}$ and $\psi(x,\vartheta,T) = h(x)$ was
shown to have a unique solution in \cite{BBP1997} when the coefficients of $\mathcal{L}$ satisfy some given Lipschitz conditions. In fact when
$(x,\vartheta,t)$ and $(x',\vartheta',t') \in \mathcal{O}^\delta$ we have
$|\sqrt{\vartheta'}-\sqrt{\vartheta}| \leq \frac{1}{2 \delta}
|\vartheta'-\vartheta|$ and so the operator $\mathcal{L}$
satisfies the assumptions made in \cite{BBP1997}. The extension of
the uniqueness result to our current setting is straightforward
and we only give the main details.

We first show that $u_1 - u_2$ is a subsolution of a related IPDE.
Suppose $\psi \in \mathcal{W}\cap C^2$ and $u_1-u_2-\psi$ attains a
maximum at $(y_0,t_0) \in \mathcal{O}^\delta$. Set
\begin{eqnarray*} w(y_1,y_2,t,s) = u_1(y_1,t)-u_2(y_2,s)\end{eqnarray*} and \begin{eqnarray*} \phi(y_1,y_2,t,s) =
 \frac{1}{2 \epsilon}
|y_1-y_2|^2 + \frac{1}{2 \alpha} |t-s|^2 +
\psi(y_1,t).\end{eqnarray*}
 Since $u_1$ and $u_2$ are in
$\mathcal{W}$, the function $w-\phi$ attains its maximum
$(y_1^*,y_2^*,t^*,s^*)$ (which depends on $\epsilon,\alpha$) in
$\mathcal{O}^\delta \times \mathcal{O}^\delta.$ By a classical argument in
the theory of viscosity solutions we can show that
 $\frac{1}{ \epsilon} |y_1^*-y_2^* |^2, \frac{1}{
\alpha} |t^*-s^*|^2 \rightarrow 0$ when $\epsilon, \alpha
\rightarrow 0$ and \begin{eqnarray*}(y_1^*,y_2^*,t^*,s^*)
\rightarrow (y_0,y_0,t_0,t_0)\end{eqnarray*} when $\epsilon,
\alpha \rightarrow 0.$

Applying Theorem 8.3 of Crandall et al. \cite{CIL1992} to the
functions $w$ and $\phi$, we find matrices $Y_1,Y_2$ such that
\begin{eqnarray*} \Big(a+\frac{\partial \psi}{\partial
t}(y_1^*,t^*), b + D \psi (y_1^*,t^*), Y_1 \Big)
&\in& \overline{\mathcal{J}}^{2,+} u_1(y_1^*,t^*)\\
\Big(-a,-b, -Y_2 \Big) &\in& \overline{\mathcal{J}}^{2,+}
(-u_1)(y_2^*,s^*).
\end{eqnarray*}
with $a= \frac{1}{\alpha}(t^*-s^*)$ and $b =
\frac{1}{\epsilon}(y_1^* - y_2^*)$ and for $0<\xi<1$ the
inequalities \begin{eqnarray*}\max \Big(a +
\mathcal{L}_\xi^{b,Y_1}[u_1,\psi](y_1^*,t^*)+\frac{\partial
\psi}{\partial t}(y_1^*,t^*)+\volp_1^*\frac{\partial^2
\psi}{\partial x^2}(y_1^*,t^*) - r u_1(y_1^*,t^*) \: ; \:\\ \quad
\quad h(x_1^*)-u_1(y_1^*,t^*) \Big) \geq  0 \end{eqnarray*}  and
\begin{eqnarray*} \max \Big(a +
\mathcal{L}_\xi^{b,Y_2}[u_2,\psi](y_2^*,s^*) - r u_2(y_2^*,s^*) \: ; \:
h(x_2^*)-u_2(y_2^*,s^*) \Big) \leq  0 \end{eqnarray*} are satisfied.
Write these two expressions as $\max(A,B) \geq 0$ and $\max(C,D)
\leq 0.$ Then $\max(A-C,B-D)\geq 0$. Now,
$B-D=h(x_1^*)-u_1(y_1^*,t^*) - h(x_2^*)+u_2(y_2^*,s^*),$ and because
$h$ is Lipschitz $|h(x_1^*)-h(x_2^*)|\rightarrow 0$ when
$\epsilon, \alpha \rightarrow 0$. Thus $B-D \rightarrow
u_2(y_0,t_0)-u_1(y_0,t_0).$ On the other hand it was shown in
\cite{BBP1997} that
\begin{eqnarray*}A-C &\leq&
 r (u_2(y_2^*,s^*)-u_1(y_1^*,t^*))+\frac{1}{\epsilon} (\frac{1}{2}- \lambda + \frac{1}{4 \delta})
|y_1^*-y_2^*|^2
\\&& +
\frac{\partial \psi}{\partial t} +  (r'-\frac{1}{2} \volp_1^*,
\volp_1^*) D \psi(y_1^*,t^*) + \volp_1^* \frac{\partial^2
\psi}{\partial x^2}(y_1^*,t^*) \\&&+ \lambda \int_0^\infty
\Big(\psi(x_1^*+\rho z, \volp_1^*+
z,t^*)-\psi(y_1^*,t^*) - z  (\rho, 1) \cdot D \psi \Big) W(dz)\\
&& + \lambda \int_0^{\xi} \Big(\phi(x_1^*+\rho z,\volp_1^* +
z,x_2^*,\volp_2^*,t^*,s^*)
-\phi(x_1^*,\volp_1^*,x_2^*,\volp_2^*,t^*,s^*)
\\&& \qquad \qquad -z  (\rho,1
) \cdot (b+D \psi(y_1^*,t^*)) \Big) W(d z) \\
&& - \lambda \int_0^{\xi} \Big(\phi(x_1^*,\volp_1^*
,x_2^*+\rho z,\volp_2^*+ z,t^*,s^*)
-\phi(x_1^*,\volp_1^*,x_2^*,\volp_2^*,t^*,s^*)
\\&& \qquad \qquad -z (\rho,1
) \cdot b_2 \Big) W(d z),\end{eqnarray*} in which $r'=(r-\lambda
\kappa^y(\rho) + \lambda \mu_2).$
 Using the fact that  $\phi \in \mathcal{W} \cap \mathcal{C}^2$ we
find letting $\xi \rightarrow 0$ and then $\epsilon, \alpha
\rightarrow 0$ that
\begin{eqnarray}\label{subsolutionphi}A-C &\leq& -r (u_1(y_0,t_0)-u_2(y_0,t_0))  + \frac{\partial \psi}{\partial t} + \mathcal{L}
\psi.\end{eqnarray} Consequently,
\begin{eqnarray}\nonumber\max(-r (u_1-u_2)(y_0,t_0) +
\frac{\partial \psi}{\partial t}(y_0,t_0) +
\mathcal{L}\psi(y_0,t_0),\\ \quad \quad  -(u_1-u_2)(y_0,t_0)) \geq 0
\label{psiviscosity}.\end{eqnarray}

As shown in \cite{BBP1997} (see lemma 3.8), there exists a
function $\chi \geq 1$ such that \begin{eqnarray*}\frac{\partial
\chi}{\partial t}+ \mathcal{L} \chi - r \chi < 0\end{eqnarray*}
and for which the maximum
\begin{eqnarray*}M=\sup_{\mathbb{R}\times\mathbb{R}_+\times[t_1,T]} ((u_1-u_2)(y,t)- \beta
\chi(y,t))e^{r(t-T)}\end{eqnarray*}is attained at some point
$(y_0,t_0)$.  Then
\begin{eqnarray*}(u_1-u_2-\beta \chi)(y,t)  \leq
(u_1-u_2-\beta \chi)(y_0,t_0) e^{r(t_0-t)}.
\end{eqnarray*} Let $\psi(y,t) = \beta \chi(y,t)-(u_1-u_2- \beta \chi)(y_0,t_0)
e^{r(t_0-t)}.$ Then $\psi$ satisfies the properties in the subsolution definition, hence it satisfies Equation \ref{psiviscosity}. But
\begin{eqnarray*}
(\frac{\partial \psi}{\partial t}+ \mathcal{L} \psi) (y_0,t_0) &=&
(\beta \frac{\partial \chi}{\partial t} + r (u_1-u_2-\beta \chi)+ \beta \mathcal{L} \chi)
(y_0,t_0)\\
&<&  r(u_1-u_2)(y_0,t_0).
\end{eqnarray*} Hence, either $(u_1-u_2)(y_0,t_0) \leq 0,$ or
$\volp_0 = \delta$ or $t_0=T$ and, in this case, $(u_1-u_2)(y_0,t_0)\leq \epsilon$ by assumption.
 Hence, we
conclude that
\begin{eqnarray*}
(u_1-u_2)(y,t) &\leq&  \beta \chi(y,t)- \beta
\chi(y_0,t_0)e^{r(t_0-t)} + (u_1-u_2)(y_0,t_0) e^{r(t_0-t)} \\ &\leq& \beta \chi(y,t)+ \epsilon e^{r(t_0-t)} .
\end{eqnarray*}
Sending $\beta$ to zero we get $u_1  \leq  u_2 +\epsilon e^{r (T-t)}$ on $\mathbb{R} \times (\delta, \infty) \times [t_1,T]$. As done
in \cite{BBP1997}, we can repeat this argument as many times as
needed to get $u_1  \leq  u_2 + \epsilon e^{r (T-t)}$ on $\mathcal{O}^\delta$.
\end{proof}

A solution of (\ref{AmIPDE})-(\ref{AmIPDEboundarycond}) is said to be minimal if it is less or equal to any other solution of (\ref{AmIPDE})-(\ref{AmIPDEboundarycond}).

\begin{theorem}\label{thMinimal}
$u$ is the minimal viscosity solution of (\ref{AmIPDE})-(\ref{AmIPDEboundarycond}).
\end{theorem}

\begin{proof}
Let $\delta>0$ and define
\begin{eqnarray*}
u^\delta(x,\volp,t) = \sup_{\tau \in \mathcal{T}_{T-t}, \tau \leq \tau_\delta} \mathbf{E}\left(e^{-r \tau} h (X_\tau^{x,\volp})\right)
\end{eqnarray*} in which
$$ \tau_\delta = \inf\{s \geq 0 : \vol_s^\volp \leq \delta\}.$$ Then $u^\delta$ is a viscosity solution of (\ref{AmIPDE}) on $\mathcal{O}^\delta$ with boundary conditions
\begin{eqnarray}\label{AmIPDEboundarydelta}
u^\delta(x,\volp,t) = h(x) \mbox{ for $t=T$ or $\volp=\delta.$}\end{eqnarray} The proof of this statement is essentially the same as the proof for the viscosity solution property of $u$. The main difference is that the maturity $T$ is replaced by $\tau_\delta.$
Note that  $\vol_s^{\delta'}> \delta$  for $\delta' > \delta e^{\lambda T},$ hence $u^\delta(x,\volp,t) = u(x,\volp,t)$ for all $x \in \mathbb{R},$ $t<T$ and $\volp > \delta e^{\lambda T}.$

Let $\tilde{u}$ be another viscosity solution of (\ref{AmIPDE})-(\ref{AmIPDEboundarycond}). Then $\tilde{u}$ is a viscosity solution of (\ref{AmIPDE}) on $\mathcal{O}^\delta$ with boundary values
$\tilde{u}(x,\volp,t)$ for $t=T$ or $\volp=\delta.$ Also, $\tilde{u}(x,\volp,t) \geq h(x)= u^\delta(x,\volp,t)$ for $t=T$ or $\volp=\delta.$ By Theorem \ref{thComparison}, we find that $\tilde{u} \geq u^\delta$ on $\mathcal{O}^\delta.$ In particular, $\tilde{u}(x,\volp,t) \geq u(x,\volp,t)$ for $x \in \mathbb{R},$ $t<T$ and $\volp > \delta e^{\lambda T}.$ Since $\delta$ is arbitrary, $\tilde{u} \geq u$ on $\mathcal{O}.$
\end{proof}

Following Pham \cite{P1998}, we denote by $UC_{x,\volp}(\mathcal{O})$ the set of functions defined on $\mathcal{O}$ uniformly continuous in $(x,\volp)$, uniformly in $t$. We have already shown that the function $u$ satisfies
\begin{eqnarray*}
|u(x',\volp',t)-u(x,\volp,t)| \leq C\left(|x'-x|+|\volp'-\volp|+\sqrt{|\volp'-\volp|}\right).
\end{eqnarray*} Hence, $u \in UC_{x,\volp}(\mathcal{O})$. Using the two previous theorems, we can show the uniqueness in $UC_{x,\volp}(\mathcal{O})$.

\begin{theorem}
$u$ is the unique viscosity solution of (\ref{AmIPDE})-(\ref{AmIPDEboundarycond}) in $UC_{x,\volp}(\mathcal{O})$.
\end{theorem}

\begin{proof} Let $\tilde{u} \in UC_{x,\volp}(\mathcal{O})$ be another viscosity solution of (\ref{AmIPDE})-(\ref{AmIPDEboundarycond}).
Let $\epsilon >0.$ Then there exists $\delta>0$ such that $0 \leq u(x,\volp,t) - u(x,0,t) = u(x,\volp,t) - h(x) < \epsilon$ and  $0 \leq \tilde{u}(x,\volp,t) - \tilde{u}(x,0,t) = \tilde{u}(x,\volp,t) - h(x) <  \epsilon$ for $\volp \leq \delta.$ In particular, $|u(x,\volp,t)- \tilde{u}(x, \volp,t)| < \epsilon$ for all $x$, all $t$ and $\volp \leq \delta.$ Furthermore, by Theorem \ref{thMinimal},  we obtain that $u(x, \delta,t) \leq \tilde{u}(x,\delta,t)  \leq u(x, \delta,t)+ \epsilon,$ and $u(x, \volp,T) = \tilde{u}(x,\volp,T)$ by definition. By the comparison principle of Theorem \ref{thComparison}, we find that  $u(x, \volp,t) \leq \tilde{u}(x,\volp,t)  \leq u(x, \volp,t)+ \epsilon e^{r (T-t)}$ for all $(x,\volp,t) \in \mathcal{O}^\delta.$ Hence,
$u(x, \volp,t) \leq \tilde{u}(x,\volp,t)  \leq u(x, \volp,t)+ \epsilon e^{r T}$ for all $(x,\volp,t) \in \mathcal{O}.$ Since $\epsilon$ is arbitrary, we obtain the desired result.
\end{proof}

\par\bigskip\noindent
{\bf Acknowledgment.} Some parts of this paper were done at HEC Montréal and the author
would like to thank Bruno Rémillard and an anonymous referee for
their help. This work was supported in part by the NSERC, the IFM2
and the Fonds québécois de la recherche sur la nature et les
technologies.

\bibliographystyle{amsplain}

\end{document}